\input{aipcheck}

\documentclass[
    ,final            
  ]
  {aipproc}

\layoutstyle{6x9}

\begin{document}

\vspace*{-1cm}
\title{Variable precession of the NS in Her X-1}


\author{R. Staubert}{
  address={Institut f\"ur Astronomie und Astrophysik, Universit\"at T\"ubingen, Germany}
}

\author{D. Klochkov}{
  address={Institut f\"ur Astronomie und Astrophysik, Universit\"at T\"ubingen, Germany}
}

\author{K. Postnov}{
  address={Sternberg Astronomical Institute, Moscow, Russia}
}

\author{N. Shakura}{
  address={Sternberg Astronomical Institute, Moscow, Russia}
}

\author{J. Wilms}{
  address={Dr. Remeis Sternwarte, Astronomisches Institut, Universit\"at Erlangen, Germany}
}

\author{R.E. Rothschild}{
  address={Center for Astrophysics and Space Science, University of California, San Diego, USA}
}

\vspace*{-0.3cm}
\begin{abstract}



We present evidence for an identical behavior of the precession of the accretion disk 
and that of the neutron star (NS) in Her~X-1, based on investigating the well established 
35\,day modulation in Her~X-1 in two different ways: 
1) following the turn-ons, thought to be due to the precession of the accretion disk, and 
2)  following the re-appearance of the shape of the pulse profiles, which we 
assume to be due to precession of the NS. The turn-on evolution and the evolution 
of the 'phase-zero' values of the precessing NS (as determined from the pulse profiles) 
track each other very closely. Since the turn-on evolution is strongly correlated with the 
pulse period evolution, this means that there is also a strong correlation between the 
spin and the precession of the NS. 
There is a very strong physical coupling between the 
NS and the accretion disk, we suggest through physical feed-back in the binary system.
The apparent long-term stability of the 35\,d clock may be due to the 'interior' of the NS, 
the coupling of which to the observable surface effects is of general importance for the 
physics of super-dense, highly magnetized material.
\end{abstract}

\maketitle


\vspace*{-1.2cm}
\subsection{Identical precessional behavior of neutron star and accretion disk}
\vspace*{-0.3cm}
A spinning non-spherical body will generally show precessional motion, in the torque-free 
case 'free precession'. Several radio pulsars show evidence for free
precession. For Her~X-1 systematic variations of the shape of the X-ray pulse profiles 
with phase of the 35\,day modulation has lead to the suggestion of a freely precessing 
neutron star (Tr\"umper et al. 1986, Shakura et al. 1999): the beamed emission from the 
accreting hot spot varies for a distant observer. We point out, however, that in Her~X-1
we do not have a torque-free case, so its precession is not entirely 'free'.
The 35\,day modulation of the X-ray flux is generally explained by the precession
of the accretion disk which regularly blocks the view to the X-ray emitting regions
near the magnetic poles of the neutron star. A strong correlation is known to exist 
between the histories of the spin period and that of the turn-ons (Staubert et al. 2006, 
2009). Here we show that all three physical parameters - the spin of the NS, the 
precession of the accretion disk, and the precession of the NS itself - are highly 
correlated.

Using pulse profiles in the 9-13\,keV range observed with 
\textsl{RXTE} (35\,d cycle nos. 307, 308, 313 and 323; for the definition of cycle counting 
see Staubert et al. 2009), we have produced a template with a 'mean observed profile' for 
every 0.01 in 35\,d phase. Any observed profile can then be compared to this template and 
the 35\,d phase measured (by $\chi^{2}_{min}$ fitting), generally to an accuracy of 
+/- 0.02 in phase. This phase value (or several such values for a particular Main-On 
cycle) can then be used to determine the absolute time of 'phase-zero', which we
identify with the phase-zero of the NS precession.

\newpage

\begin{figure}
  \includegraphics[height=.2\textheight]{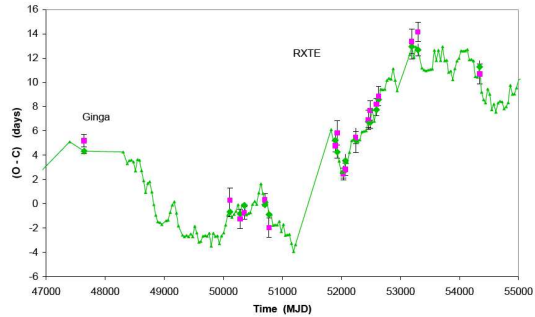}
  \caption{(O-C) values for observed turn-ons (green points and curve) and for the so far generated neutron star 'phase-zero' times from pulse profile fitting (mangenta points with uncertainties).}
 \end{figure}

\vspace*{-0.8cm}
Fig.~1 shows the (O-C) (= 'observed minus calculated') diagram for all observed 
turn-ons since the discovery of the source. 'Calculated' turn-ons are based on a 
constant period of 34.85\,days - a horizontal line in Fig.~1. Any other slope in (O-C)
defines a different precessional period. So far NS phase-zero values were produced
for most of the \textsl{RXTE} observations (1997-2005) and for a set of \textsl{Ginga}
observations (1989) and one \textsl{Integral} observation (2007), from pulse profiles
in the energy range 9-13\,keV. It is evident that the two (O-C) histories track each other
quite closely. A similar result has been obtained by fitting pulse profiles with a theoretical 
model (Postnov 2004, and ongoing work). This is not only so on long time scales 
(several years), but also on much shorter time scales (several tens of days).
The physical coupling between the NS and the accretion disk must be very
strong. Staubert et al. (2009) have proposed a physical model for such a coupling, based 
on a closed-loop feedback. \\
\indent\textsl{Conclusions.} Under the assumption that both of our working hypotheses are correct - namely that the 
flux modulation is due to the precession of the accretion disk and the variations of the 
pulse profiles are due to the precession of the NS -, we reach the following conclusions: 
1) The period of precession of the NS can change on short time scales
(a few tens of days).
2) The precession of the NS (phase-zero from pulse profile fitting) varies in the same 
way as the precession of the accretion disk. 
3) The spin of the NS is strongly correlated to the precession of the accretion disk
(Staubert et al. 2006), so the two periods of the NS - the spin and the precession -
are strongly correlated (but, as it appears, in a different way as expected from free precession).
We suggest that this is due to the torque applied to the NS
through the interaction of the magnetosphere with the inner part of the accretion disk.
We are investigating whether a change in the orientation of the axes of inertia could 
be responsible for the change in precessional period. \\
\indent\textsl{Acknowledgements.} We acknowledge support through grants: DFG Sta 173/31-2, DFG 436RUS113/717 and RFBR 09-02-00032.

\end{document}